\documentclass[twocolumn,showpacs,preprintnumbers,amsmath,amssymb,superscriptaddress]{revtex4}
\usepackage{graphicx}
\usepackage{dcolumn}
\usepackage{bm}
\usepackage{longtable}

\begin{document}

\preprint{APS/123-QED}

\title{A Hebbian approach to complex network generation}

\author{E. Agliari}
\affiliation{Dipartimento di Fisica, Universit\`a degli Studi di
Parma, viale Usberti 7/A, 43100 Parma, Italy}
\affiliation{Istituto Nazionale di Fisica Nucleare, Gruppo Collegato di
Parma}
\affiliation{Theoretische Polymerphysik, Albert-Ludwig-Universit\"{a}t, Freiburg, Germany}
\author{A. Barra}
\affiliation{Dipartimento di Fisica, Sapienza Universit\`a di
Roma, P.le A. Moro 5, 00182, Roma, Italy}
\affiliation{Gruppo Nazionale per la Fisica Matematica, Sezione di Roma1}
\date{\today}

\begin{abstract}
Through a redefinition of patterns in an Hopfield-like model, we introduce and develop an approach to model discrete systems made up of many, interacting components with inner degrees of freedom.
Our approach clarifies the intrinsic connection between the kind of interactions among components  and the emergent topology describing the system itself; also, it allows to effectively address the statistical mechanics on the resulting networks.
Indeed, a wide class of analytically treatable, weighted random graphs with a tunable level of correlation can be recovered and controlled. We especially focus on the case of imitative couplings among components endowed with similar patterns (i.e. attributes), which, as we show, naturally and without any a-priori assumption, gives rise to small-world effects. 
We also solve the thermodynamics (at a replica symmetric level) by extending the double stochastic stability technique: free energy, self consistency relations and fluctuation analysis for a picture of criticality are obtained.
\end{abstract}

 \maketitle

The performance of most complex systems, from the cell to the Internet, emerges from the collective activity of many inner components; At an abstract level, the latter can be reduced to a series of nodes that are connected each other by links, envisaging the interaction. Nodes and links together form a network \cite{reviews,science,libro}.

The view offered by network description has led to identify classes of (topological) universality \cite{chun}, and to evidence how experimentally-revealable features, e.g. cliquishness, modularity, assortativity or peculiar degree distribution, not only underlie a certain degree of correlation among components and/or links but also crucially affect the behavior of the system \cite{reviews}.
This constituted 
a real breakthrough with respect to the previous tendency to model complex networks either as regular objects, such as square or diamond lattices, or as (purely uncorrelated) random networks \`a la Erd\"os-Renyi (ER) \cite{ER}.

In the first part of this work we show that, within our approach, the kind of interaction (e.g. imitative, repulsive, etc.) among components naturally gives rise to a broad class of weighted random graphs, 
whose topological properties can be properly tuned.
Such a connection also allows to infer about the plausibility of a given modelization: 
the choice of a particular network must be consistent with the kind of interactions governing the system itself and viceversa,
ultimately reflecting the internal structure of the considered nodes.

In the second part of this work, we study the thermodynamic properties of a subset of these structures generated by a ``Hebbian-like kernel''; interestingly, such an approach allows to work out the thermodynamics of a wide class of diluted graphs, even in the presence of ferromagnetic disorder on couplings.

\emph{Modelization.} 
Given a set of $V$ components, each characterized by a ``pattern'' $\xi$ (similarly to the ``hidden variable'' approach \cite{caldarelli,pastor}) drawn from a given probability distribution, a pair of nodes $i$ and $j$ is linked according to a proper rule $r(\xi_i,\xi_j)$. More precisely, here, $\xi$ is a vector given by binary string of length $L$, which encodes a set of attributes characterizing each node;
Then, 
the function $r$
associates each couple of strings to a real value which provides the pertaining coupling: $r(\xi_i,\xi_j) = J_{ij}$.

Now, crucial for the whole approach are the way the strings are generated and the rule $r$, both to be defined according to the processes one wishes to model.

\emph{The Hebbian-like kernel} 
We investigate in detail the case of biased patterns where the probability to extract any entry is $P(\xi_i^{\mu} = 0) = 1 - P(\xi_i^{\mu} = 1) = (1-a)/2, \; a \in [-1,1]$; the rule $r$ given by the scalar product among strings
 \begin{equation} \label{eq:J_ij}
J_{ij}=J_{ji}=
\sum_{\mu=1}^L \xi_i^{\mu}\xi_j^{\mu}.
\end{equation}
Notice that such a rule resembles the Hebbian kernel well-known in neural networks \cite{amit}, apart from the shift $[-1, +1] \to [0,+1]$ in the definition of patterns; this plain replacement converts frustration into ferromagnetic dilution. 
We are therefore focusing on systems where the interaction among components is stronger the larger the number of attributes they share.

\begin{figure}[tb] \begin{center}
\includegraphics[width=.48\textwidth]{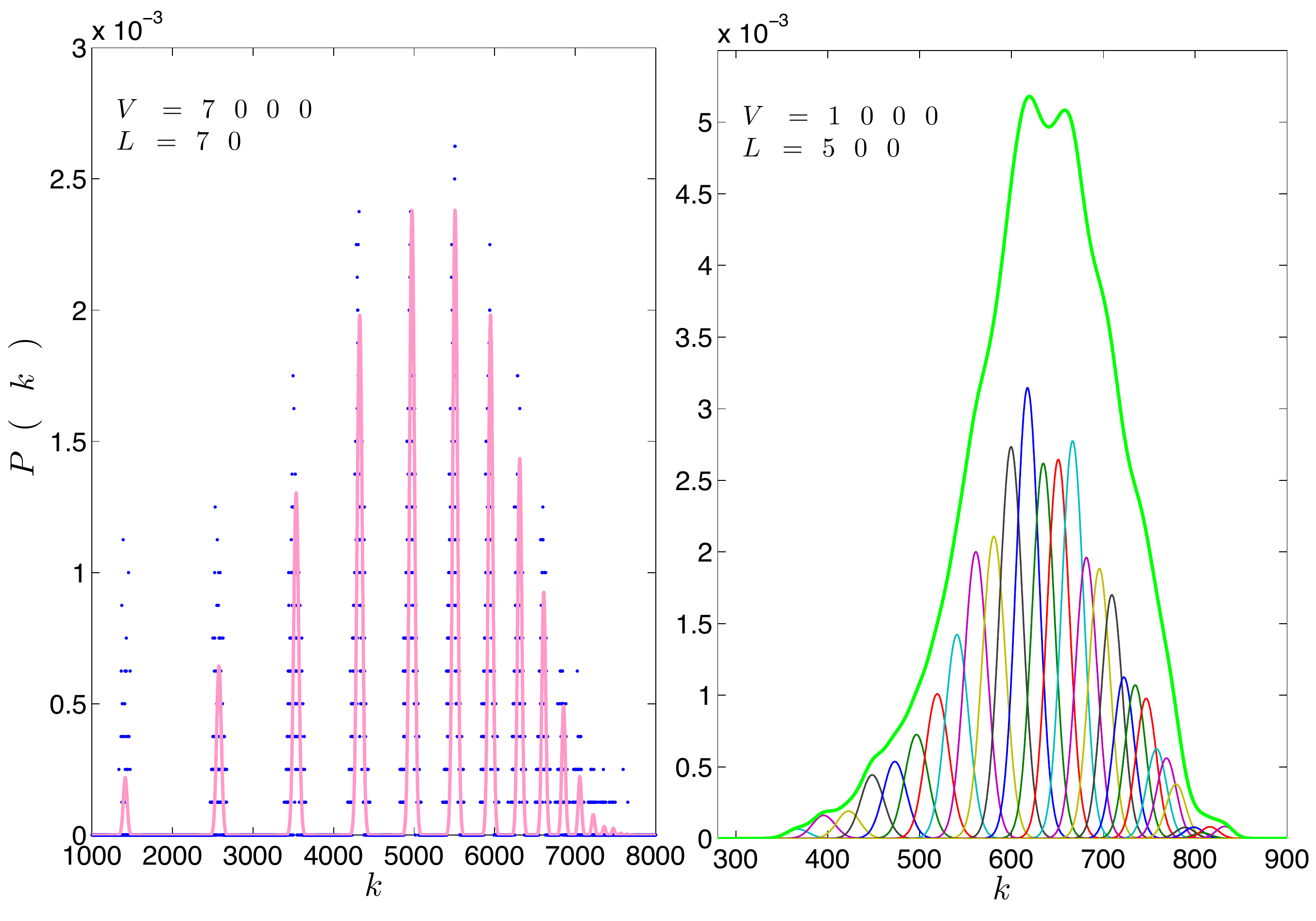}
\caption{\label{fig:distr} (Color on line) Degree distribution $\bar{P}_{\mathrm{degree}}(z;a,L,V)$ for systems displaying small ratio $L/V$ and multimodal distribution (left panel) and large $L/V$ with modes collapsing into a unimodal distribution. In the former case we plot data points ($\bullet$) as well as the the analytical curve provided by Eq.~\ref{eq:overall_degree}. In the latter case we show the distributions corresponding to each mode $P_{\mathrm{degree}}(z;a,\rho,V)$ in agreement with numerical data, as well as the overall distribution (thicker, bright curve).}
\end{center}
\end{figure}

An important parameter characterizing a given node, is the number $\rho$ of non-null entries present in the related string: Due to the independence underlying the extraction of each entry, $\rho$ follows a binomial distribution $P_1(\rho;a,L) = \mathcal{B}(\rho; L, (1+a)/2)$ with average $\bar{\rho}_{a,L} = L (1+a)/2$.
Moreover, it can be shown that the probability for two string $\xi_i$ and $\xi_j$, displaying respectively $\rho_i$ and $\rho_j$ non-null entries, to be connected is $P_{\mathrm{link}}(\rho_i,\rho_j; L) = 1 - (L- \rho_i)! (L-\rho_j)! / [L! (L - \rho_i - \rho_j)!]$.
By averaging over, say $\rho_j$, one finds the expected link probability $P_{\mathrm{link}}(\rho_i, a) = 1 - [(1-a)/2]^{\rho_i} $ for the node $i$. Then, the probability that $i$ has degree (number of neighbors) equal to $z$ follows as the binomial $P_{\mathrm{degree}}(z; a,\rho_i, V) = \mathcal{B}(z;V,P_{\mathrm{link}}(\rho_i, a) )$.
Due to the average over $\rho_j$, such a degree distribution corresponds to a mean-field approach where we treat all the remaining nodes in the average; this works finely for $V$ large and $\rho_i$ not too small (see also Fig.~\ref{fig:distr}).
Therefore, the number of null-entries  controls the
degree distribution of the pertaining node: A large $\rho$ gives
rise to narrow (small variance) distributions peaked at large
values of $z$.
Then, the global degree distribution reads off as a combination of binomials
\begin{equation} \label{eq:overall_degree}
\bar{P}_{\mathrm{degree}}(z;a,L,V) = \sum_{\rho=0}^{L}
P_{\mathrm{degree}}(z;a,\rho,V)P_1(\rho;a,L),
\end{equation}
giving rise to a $L$-modal distribution, where ``modes'', each corresponding to a different value of $\rho$, are solved as long as the connectivity and $L$ are not too large in order to ensure spread distributions for $z$ and $\rho$ (see Fig.~{\ref{fig:distr}, left and right panels, respectively and \cite{nostro} for more details).

Multimodal degree distributions constitute an interesting feature of the model as they allow to naturally discriminate between different classes of nodes possibly fulfilling different functions. In particular, nodes corresponding to a large degree are often associated to a relatively small reactivity and vice versa \cite{immuno,palla}.

A more global characterization can be attained by the average link probability, applying to a generic couple of nodes, neglecting any information about correlations
\begin{equation} \label{eq:giusto}
p = 1 - \left[ 1 - \left( \frac{1+a}{2} \right)^2 \right]^L,
\end{equation}
so that the average degree reads as $\bar{z} = p V$.
Now, for $L \rightarrow \infty$, $p$ approaches a
discontinuous function assuming value $1$ when $a>-1$ and value
$0$ when $a=-1$. To fix ideas, let us focus on the the so-called high-storage regime where $L$ is
linearly divergent with $V$, i.e. $L=\alpha V$, then, the range of values for $a$ yielding a
non-trivial topology can be characterized by means of the scaling
\begin{equation}\label{eq:scaling}
a = -1 + \frac{\gamma}{V^{\theta}},
\end{equation}
where $\theta \geq 0$ and $\gamma$ is a finite parameter \footnote{Since $a \in [-1, 1]$, we have that $0 \leq \gamma \leq 2 V^{\theta}$; in particular, when $\theta=0$ the upper bound for $\gamma$ is $2$.}.

Following \cite{nostro}, it is possible to distinguish the following regimes (see Fig.~\ref{fig:regimes}):
\begin{itemize}
\item  $\theta <1/2$, \; $p \approx 1$, \; $\bar{z} \approx V$  \; $\Rightarrow$  Fully connected graph\\
\item  $\theta =1/2$, \; $p \sim 1- e^{-\gamma^2 \alpha/4} \sim \gamma^2 \alpha/4$, \;  $\bar{z} = O(V)$ \; $\Rightarrow$ Linearly diverging connectivity; Within a mean-field description the ER random graph with finite probability is recovered.
\item  $1/2 < \theta <1$, \; $p \sim \gamma^2  \alpha V^{1-2 \theta}/4$, \; $\bar{z} = O(V^{2 - 2 \theta})$ \; $\Rightarrow$  Extreme dilution regime: $\lim_{V \rightarrow \infty}  \bar{z}^{-1} = \lim_{V \rightarrow \infty} \bar{z}/V = 0$ \cite{watkin}.\\
\item  $ \theta = 1$, \; $p \sim \frac{\gamma^2  \alpha}{4V}$, \; $\bar{z} = O(V^{0})$ \; $\Rightarrow$  Finite connectivity regime; Within a mean-field description $\gamma^2 \alpha / 4 = 1$ corresponds to a percolation threshold.
\end{itemize}
Larger values of $\theta$ determine a disconnected graph with vanishing average degree.

Therefore, while $\theta$ controls the connectivity regime of the network, $\gamma$ allows a fine tuning. 

\begin{figure}[tb] \begin{center}
\includegraphics[width=.48\textwidth]{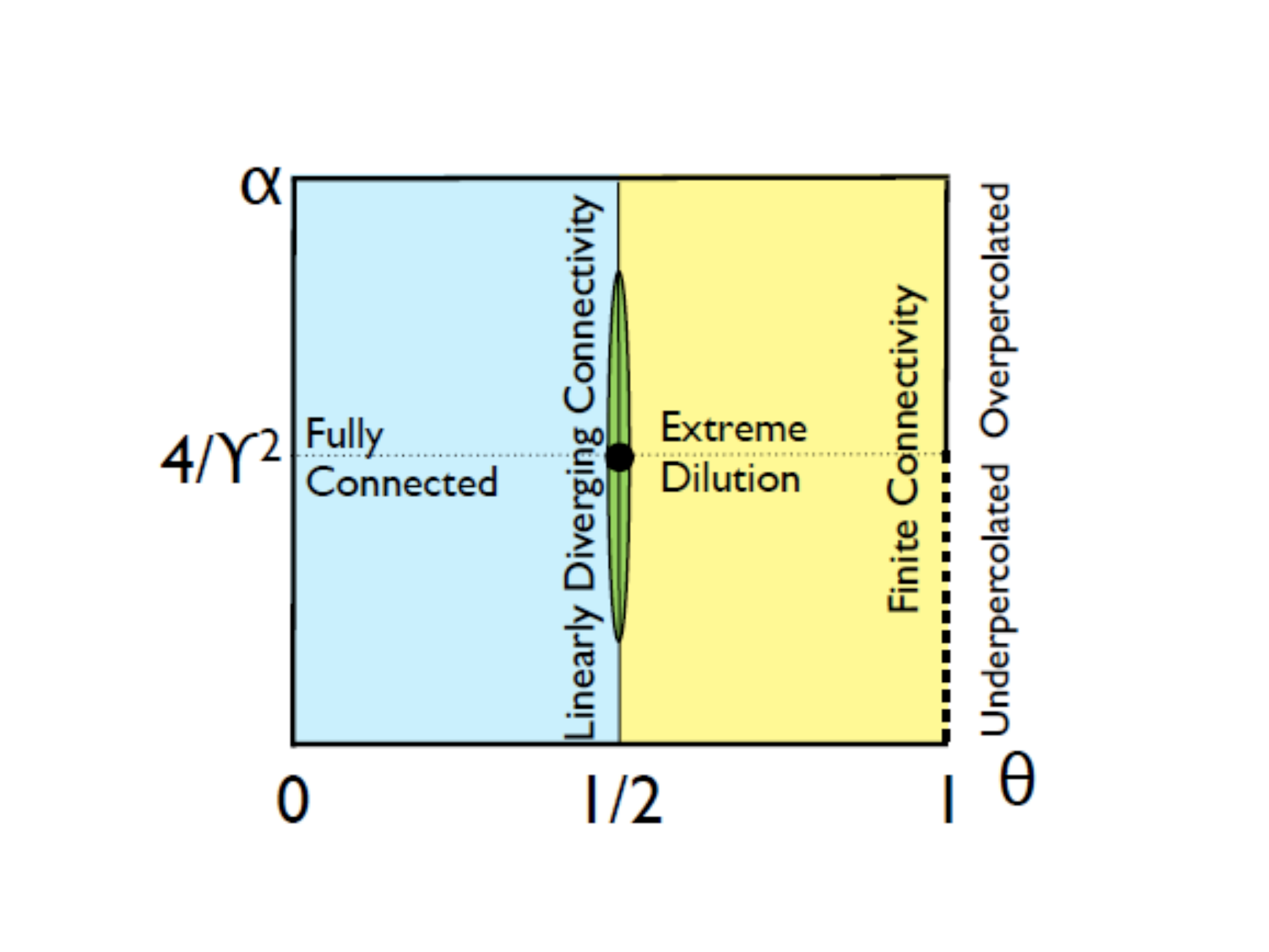}
\caption{\label{fig:regimes} (Color on line) Phase diagram representing the topology of the graph as the parameters $\theta$, $\gamma$ and $\alpha$ are varied. In the region on the left (blue) disorder on couplings is still present, while on the right side (yellow) disorder on couplings is lost while topological inhomogeneity is still present. In the narrow central region (green) a coexistence of  both kinds of disorder is achieved.}
\end{center}
\end{figure}

Up to now we just focused on topological disorder; as for couplings, we can still detect ``modes'', each characterized by $J_{\rho}$ representing the average strength for links stemming from a node  associated to a string with $\rho$ non-null entries. While $J_{\rho}$ provides a measure of the local ``field'' seen by a single node, a global description can be attained by the overall average coupling $\bar{J} \equiv \sum_{\rho} J_{\rho} P_1(\rho; a,L)$, taken over the whole graph.
The two quantities are related via $J_{\rho} = \rho/L \sqrt{\bar{J}}$ \cite{nostro} and, as we will see,
despite the
self-consistence relation (see Eq.$10$), more sensible to local
conditions, is influenced by $\sqrt{\bar{J}}$, the critical
behavior occurs at $\beta_c = \bar{J}^{-1}$ consistently with
a manifestation of a collective, global effect.

By looking in more detail at the coupling distribution holding in the thermodynamic limit and for values of $a$ determined by Eq.~\ref{eq:scaling}, we find that for $1/2 < \theta \leq 1$, nodes are pairwise either non-connected or connected due to one single matching among the relevant strings, namely, neglecting higher order corrections, $J=0$ with probability $p_0 \sim \exp(\alpha \gamma^2 V^{1 - 2 \theta}/4) \sim 1- \gamma^2 \alpha / (4 V^{2 \theta -1})$, while $J=L^{-1}$ with probability $p_1 \sim p_0 \gamma^2 \alpha / (4 V^{2 \theta -1}) \sim 1 - p_0 $. For $\theta =1/2$ this still holds when $\alpha \gamma^2 /4 \ll 1$, which corresponds to a relatively high dilution regime, otherwise some degree of disorder is maintained, being that $p_k \sim (\alpha \gamma^2/4)^k / k!$. On the other hand, for $\theta < 1/2$, while topological disorder is lost, disorder on couplings is still present. However, for $\theta = 0$ and $\gamma = 2$, the coupling distribution gets peaked at $J=1$ and, again, disorder on couplings is lost so that a pure Curie-Weiss model is recovered (see  also \cite{nostro}).

Before concluding this part, we stress that slight variations in the rule provided by Eq.~\ref{eq:J_ij} can yield dramatic changes in the global layout of the graph, e.g. scale-free degree and/or coupling distributions.

\emph{Small-World properties}  A ``small-world'' network \cite{watts} displays, by definition, diameter growing as $\log N$, hence comparably with the case of ER random graphs, and high cliqueshness indicating a high level of redundancy.

The cohesiveness of the neighbourhood of a node $i$ is usually quantified by the (local) clustering coefficient $c_i$, defined as the fraction of permitted edges between nodes adjacent to $i$ that actually exist. 
The average clustering coefficient $c = \sum_{i=1}^V c_i$ can be compared with
the one expect for a comparable (i.e. displaying same average degree) ER graph, namely $c^{\mathrm{ER}} = \bar{z}/V$.

For the graph generated by Eq.~\ref{eq:J_ij}, the neighborhood $V_i$ of $i$ is made up of all nodes displaying at least one non-null entry corresponding to any non-null entries of $\xi_i$. This condition biases the distribution of strings relevant to nodes $ \in V_i$, so that they are more likely to be connected with each other. This is also confirmed numerically: Fig.~\ref{fig:clustering} shows that $c / c^{\mathrm{ER}} >1$ in a wide region of the plane $(\alpha,a)$, and, in particular, in the region of high dilution.

\begin{figure}[b] \begin{center}
\includegraphics[width=.48\textwidth]{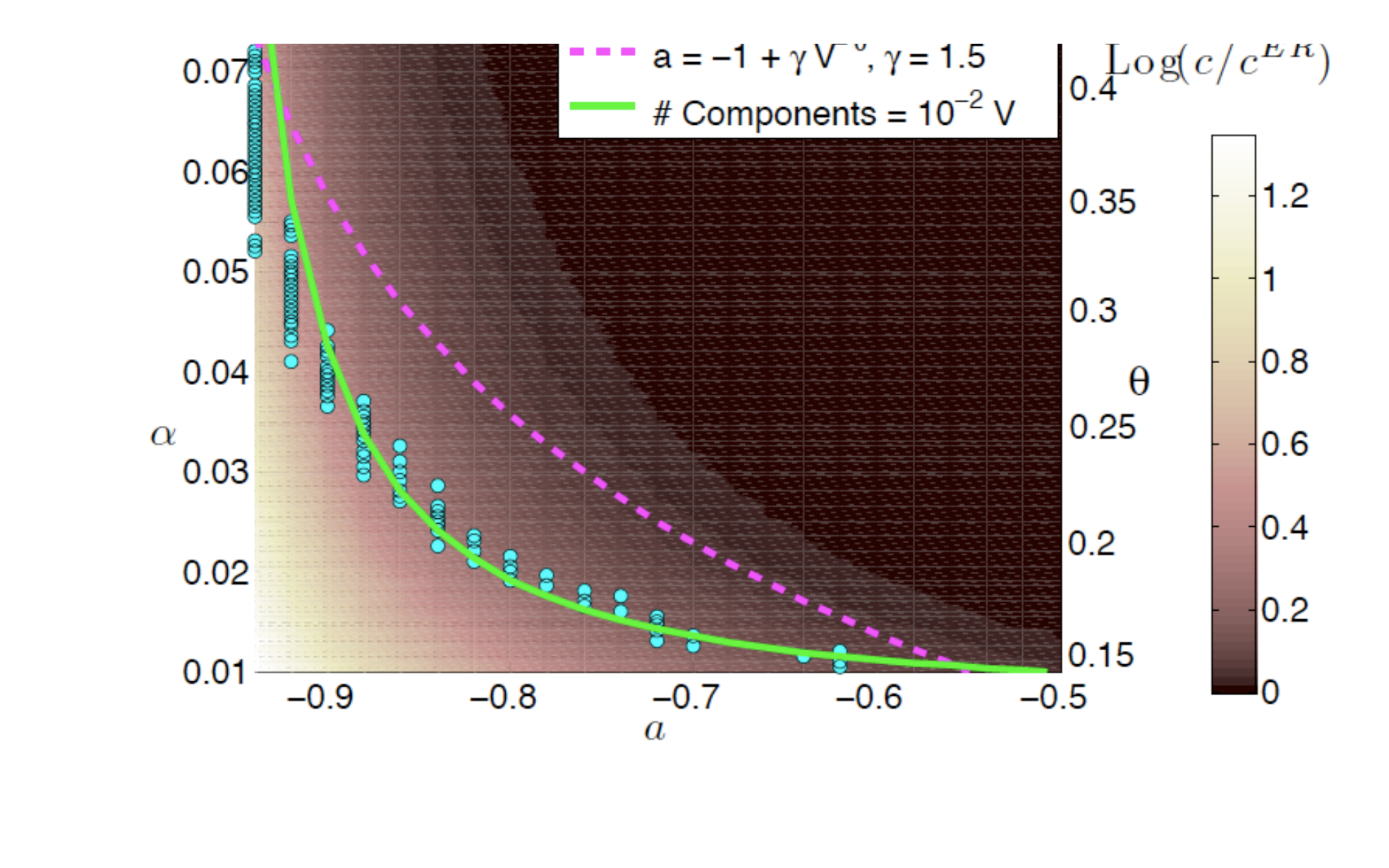}
\caption{\label{fig:clustering} (Color on line) Natural logarithm for the ratio $c/c^{\mathrm{ER}}$ as a function of $\alpha$ and $a$ for a system with $V=2000$. When $a$ follows the scaling of Eq.~\ref{eq:scaling} with $\gamma = 1.5$ (dashed line) the ratio is approximately $4$, meaning large clustering. Data points ($\bullet$), fitted by a power law $\sim (1+a)^{\eta}, \eta \approx 1.5$, demarcate the region where the graph is made up of $20$ components (a macroscopic one plus isolated nodes).}
\end{center}
\end{figure}

Interestingly, this kind of link correlation allows to detect, within the graph, communities densely and strongly linked up, so that the role of weak ties is crucial in bridging such communities and maintaing the graph connected \cite{grano,forth}.


 \emph{Thermodynamics}
 When dealing with thermodynamical properties of these networks, we first paste $V$ variables $\sigma$ on the nodes and
define the Hamiltonian
\begin{equation}\label{hamilta}
H(\sigma;\xi)=
\frac{-1}{2\alpha N^{2(1-\theta)}}\sum_{i<j}^V\sum_{\mu}^P \xi^{\mu}_i\xi^{\mu}_j \sigma_i
\sigma_j,
\end{equation}
formally identical to an Hopfield model, apart the normalization which accounts for not symmetrical w.r.t. zero quenched variables. 
Via Eq.~\ref{hamilta} the standard package of disordered statistical
mechanics can be introduced: the partition function $Z_V(\beta;\xi) =
\sum_{\sigma}\exp[-\beta H_V(\sigma; \xi)]$, the Boltzmann state
$\omega(.)=\sum_{\sigma}. e^{-\beta
H_V(\sigma;\xi)}/Z_V(\beta;\xi)$, and the related free energy as
$A(\beta,\gamma,\theta) = \lim_{V\to\infty} V^{-1} \mathbb{E} \log
Z_V(\beta;\xi),$ where $\mathbb{E}$ averages over the quenched
distributions of the bit strings $\xi$. To obtain an explicit expression of the free energy, at first, trough the Hubbard-Stratonovick transform we map the partition function of our model into a bipartite ER one, whose
parties are the $V$ Ising spins and $L$ auxiliary Gaussian fields
$z$ as
\begin{equation}
Z_V(\beta;\xi) =
\sum_{\sigma}\int_{-\infty}^{+\infty}\prod_{\mu=1}^L d\mu(z_{\mu})
e^{\frac{\sqrt{\beta/\alpha}}{N^{1-\theta}}\sum_{i,\mu}^{N,P}\xi_{i,\mu}\sigma_i z_{\mu}}, \nonumber
\end{equation}
then we enlarge the
technique of the double stochastic stability \cite{BGG2} by introducing the following interpolating
structure
\begin{eqnarray}\label{interpolante} && A(t) =  \frac{\mathbb{E}}{V} \log
\sum_{\sigma}\int_{-\infty}^{+\infty} \prod_{\mu}^L d\mu(z_{\mu}) 
 \exp[ t \frac{\sqrt{\beta/\alpha}}{V^{1-\theta}} \sum_{i,\mu} \\ && \xi_{i \mu}\sigma_i z_{\mu} +(1-t) ( \sum_{l_c=1}^L b_{l_c}
\sum_i^V \eta_i \sigma_i + \sum_{l_b=1}^V c_{l_b} \sum_{\mu}^L
\chi_{\mu} z_{\mu})], \nonumber
\end{eqnarray} where now
$\mathbb{E}=\mathbb{E}_{\xi}\mathbb{E}_{\eta}\mathbb{E}_{\chi}$,  $b_{l_c}$ [with $l_c \in(1,...,L)$], and  $c_{l_b}$ [with $l_b
\in(1,...,V)$]  are real numbers (possibly functions of
$\beta,\gamma,\theta$) to be set a posteriori.
\newline
Note that $A(t=1)$ is our goal, while $A(t=0)$ is straightforward
as it involves only one-body calculations. So we want to use the
fundamental theorem of calculus to get a sum rule, as $A(1) = A(0)
+ \int_0^1 [\partial A (t') / \partial t' ]_{t'=t}dt$, which
ultimately implies the evaluation of the $t$-streaming of eq.
(\ref{interpolante}). 
Now, we need to sort out our order parameters: as replica symmetry (RS) is expected to be conserved in ferromagnetic diluted networks, we (naturally) avoid (multi)-overlaps by defining $M_{l_b}=V^{-1}\sum_i \omega_{l_b+1}(\sigma_i)$ (and analogously for the other party trough $N_{l_c}$), where the index in $\omega$ means that we are considering all the possible magnetizations built trough only $l_b+1$ links inside the graph (as the graph is no longer weighted,  it is a microcanonical decomposition of the observable in subclusters close to the one introduced in \cite{BG1}).
Once introduced the averaged order parameters as $\langle M \rangle =
\sum_{l_b}P(l_b) M_{l_b}$, and  $\langle N
\rangle = \sum_{l_c}P(l_c) N_{l_c}$, we find that $\partial_t A(t) = S(t) - \sqrt{\beta}\gamma / 2 \sum_{l_b}\sum_{l_c}P(l_b)P(l_c)\bar{M}_{l_b}\bar{N}_{l_c}$; where the fluctuation source $S(t)$ is proportional to $\langle
( M - \bar{M})( N - \bar{N}) \rangle$ and can be
neglected at the RS level (whose order parameter
values are denoted via a bar, namely $\bar{M},\bar{N}$).  
\newline
As a consequence it is possible to solve for the RS free energy to
get 
\begin{eqnarray}\label{soloM}\nonumber
 A(\beta,\gamma, \theta) &=& \log 2 + \frac{\gamma}{2V^{\theta}}\langle \log\cosh(\sqrt{\beta}\bar{N}V^{\theta})\rangle+ \\
\nonumber &+& \frac{\beta \gamma^2}{8}\langle \bar{M} \rangle ^2-\frac{\sqrt{\beta}\gamma}{2}\langle \bar{M} \rangle \langle \bar{N} \rangle.
\end{eqnarray}
Trough the self consistent relation $\langle \bar N \rangle = \frac{\sqrt{\beta}\gamma}{2}\langle \bar M \rangle$, we can express the whole theory only via $\langle M \rangle$ (as expected looking at the original Hamiltonian (\ref{hamilta})) and we start discussing the various case of interest:
\begin{itemize}

\item $\theta=0$: Fully connected weighted regime\\
The case $\theta=0$ reduces to the Curie-Weiss model and in particular, the upper bound on $\gamma$ (i.e. $\gamma=2$) describes the un-weighted fully connected topology. Choerently its termodynamics turns out to be
\begin{eqnarray}
\nonumber
 A(\beta,\gamma &=& 2,\theta=0) = \log 2 + \frac{\gamma}{2}\log \cosh(\frac{\beta \gamma}{2}\bar{M}) \\ &-&\frac{\beta \gamma^2}{8}\bar{M}^2, \ \ \ \ \bar{M} = \tanh(\frac{\beta \gamma}{2}\bar{M}).
\end{eqnarray}
Note that, as there is only one possible network built with all the links, all the subgraph magnetizations collapse into only one, namely $P(\bar{M})= \delta(\bar{M}- M_{CW})$, where with $M_{CW}$ we mean the standard Curie-Weiss magnetization. Furthermore, note that for $\gamma = 2$ we recover exactly the CW thermodynamics.

\item $\theta = 1/2$: Standard dilution and ER regime\\
With a scheme perfectly analogous to the previous one we can write down the free energy and its related self-consistent equation as

\begin{eqnarray}
&& A(\beta,\gamma,\theta=1/2) = \log 2 + \lim_{V\to\infty}(\frac{\gamma}{2\sqrt{V}} \cdot \\ && \log\cosh(\frac{\beta\gamma}{2}\sqrt{V}\langle \bar{M}\rangle)-\frac{\beta \gamma^2}{8}\langle\bar{M}\rangle^2 ),\nonumber \\
&& \langle \bar{M} \rangle = \lim_{V \to \infty}\tanh(\frac{\beta \gamma}{2}\sqrt{V}\langle \bar{M} \rangle). 
\end{eqnarray}
Ficticious diverging contributions emerge in the thermodynamic limit because the normalization we chose for the Hamiltonian (\ref{hamilta}) gives the correct extensivity for the fully connected case only, while for $\theta=1/2$ it is $O(V^{-1})$ on an ER graph. Note in fact that the argument of  the logarithm of the hyperbolic cosine can be read as $\beta \sqrt{J}V\langle \bar{M}\rangle$.  As in standard approaches \cite{ABC} it is enough to renormalize the coupling by scaling \cite{guerra} it with the amount of nearest neighbours (that in the ER dilution scales exactly linearly with the volume size).

\item $\theta>1/2$: Extremely dilution regime and all the other cases of interest\\
With a scheme analogous to the previous one it is possible to write down the pertaining free energy, coupled with its self-consistent relation for its extremization.
\end{itemize}

Interestingly, the effective field felt by the trial spin -obtained extremizing
eq. (\ref{soloM}) with respect to $\langle M \rangle$- scales as
the square root of the coupling strength instead of a linear
behavior (this effect disappears approaching the CW limit where $\sqrt{J}\to \bar{J}$). As the interaction matrix in the Hamiltonian has been normalized (i.e. is bounded by one),
$\sqrt{\bar{J}}>\bar{J}$, the local field is ``higher with respect
to a naive expectation" \footnote{In the Curie-Weiss model critical behavior and self-consistency are both linear functions of the coupling, because there is no difference among local and global environments.}. This feature, which
is a consequence of the diverging variance 
in the bitstring distribution \cite{nostro}, however does not affect the
transition line that scales consistently
with a manifestation of a collective, global effect, as $\beta_c
\propto \bar{J}^{-1}$.
\newline
In order to prove the last statement, we have to
tackle the control of the fluctuations of the rescaled order
parameters  $\langle \mathcal{M} \rangle = \sqrt{V}\langle M-
\bar{M}\rangle$, $\langle \mathcal{N} \rangle = \sqrt{L}\langle N
- \bar{N}\rangle$: At first we need to derive the $t$-streaming of
the squares of these objects (i.e. $\langle \mathcal{M}^2
\rangle,\langle \mathcal{M}\mathcal{N} \rangle,\langle
\mathcal{N}^2 \rangle)$,  which lead to a system of coupled
ordinary differential equations in $t$, namely
\begin{eqnarray}
\langle \dot{\mathcal{M}^2} \rangle &=& 2 \langle \mathcal{M}^2 \rangle \langle \mathcal{M}\mathcal{N} \rangle, \\
\langle \dot{\mathcal{M}\mathcal{N}} \rangle &=&  \langle \mathcal{M}^2 \rangle \langle \mathcal{N}^2 \rangle + \langle (\mathcal{M}\mathcal{N})^2 \rangle, \\
\langle \dot{\mathcal{N}^2} \rangle &=& 2 \langle \mathcal{N}^2 \rangle \langle \mathcal{M}\mathcal{N} \rangle,
\end{eqnarray} 
whose Cauchy condition at
$t=0$ can be obtained immediately, such that we can solve the system,
evaluate it at $t=1$ (the original statistical mechanics
framework) and check where these fluctuations do diverge
(identifying the expected second order phase transition). By using
Wick theorem to express four point correlations trough series of
two point ones and due to internal symmetries reflecting the mean
field interactions among the two parties \cite{nostro}
the plan is fully solvable and all these fluctuations (and the
correlation $\langle \mathcal{M}\mathcal{N}\rangle$) are found to diverge
on the same line: $\beta_c= \bar{J}^{-1}$, as
intuitively expected.

\vspace{0.5cm}

\vspace{1.5cm}
This work is supported by the FIRB grant: $RBFR08EKEV$

\end{document}